\documentclass[fleqn,usenatbib]{mnras}

\usepackage{newtxtext, newtxmath, nicefrac}
\usepackage[T1]{fontenc}

\DeclareRobustCommand{\VAN}[3]{#2}
\let\VANthebibliography\thebibliography
\def\thebibliography{\DeclareRobustCommand{\VAN}[3]{##3}\VANthebibliography}

\usepackage{graphicx}
\usepackage{amsmath}
\usepackage{xcolor}

\newcommand{\txa}{{\text{a}}}

\newcommand{\txd}{{\text{d}}}

\newcommand{\calE}{{\cal{E}}}

\newcommand{\rT}{r_{\scriptscriptstyle{\text{T}}}}

\newcommand{\calET}{\calE_{\scriptscriptstyle{\text{T}}}}

\defcitealias{2022MNRAS.512.2266B}{Paper~I}
\defcitealias{2023MNRAS.519.6065B}{Paper~II}
\defcitealias{2023MNRAS.525.1795B}{Paper~III}

\title[Dynamical models with a finite extent~IV]{Self-consistent dynamical models with a finite extent -- IV. Wendland models based on compactly supported radial basis functions}

\author[M. Baes]{Maarten Baes
\\%
Sterrenkundig Observatorium, Universiteit Gent, Krijgslaan 281 S9, 9000 Gent, Belgium
}

\date{Accepted 2024 June 14. Received 2024 June 14; in original form 2024 April 2}

\pubyear{2024}

\begin{document}
\label{firstpage}
\pagerange{\pageref{firstpage}--\pageref{lastpage}}
\maketitle

\begin{abstract}
We present a new step in our systematic effort to develop self-consistent dynamical models with a finite radial extent. The focus is on models with simple analytical density profiles allowing for analytical calculations of many dynamical properties. In this paper, we introduce a family of models, termed Wendland models, based on compactly supported radial basis functions. The family of models is characterised by a parameter $k$ that controls the smoothness of the transition at the truncation radius. In the limit $k\to\infty$, the Wendland model reduces to a non-truncated model with a Gaussian density profile. For each Wendland model, the density, mass and gravitational potential are simple truncated polynomial functions of radius. Via the {\tt{SpheCow}} tool we demonstrate that all Wendland models can be supported by isotropic distribution functions. Surprisingly, the isotropic distribution function exhibits varied behaviour across different Wendland models. Additionally, each model can be supported by a continuum of Osipkov--Merritt orbital structures, ranging from radially anisotropic to completely tangential at the truncation radius. To the best of our knowledge, the Wendland models presented here are the first family of models accommodating both radial and tangential Osipkov--Merritt distribution functions. Using linear superposition, these models can easily be combined to generate Wendland models with even more diverse orbital structures. While the Wendland models are not fully representative of real dynamical systems due to their Gaussian-like density profile, this study lays important groundwork for constructing more realistic models with truncated density profiles that can be supported by a range of orbital structures.
\end{abstract}


\section{Introduction}

Even in the current era of high-performance computing and detailed multi-dimensional numerical modelling, simple analytical dynamical models remain an important tool in galaxy dynamics. Such models serve as a well-tested environment to test novel modelling approaches or analysis techniques, and they are usually the starting point for full-scale numerical simulations. 

Most of the simple dynamical models presented in the literature have an infinite extent, that is, their density is strictly positive over the entire space. Popular examples with that characteristic include the Plummer sphere \citep{1911MNRAS..71..460P, 1987MNRAS.224...13D}, the Jaffe model \citep{1983MNRAS.202..995J}, the Hernquist model \citep{1990ApJ...356..359H, 2002A&A...393..485B}, the $\gamma$-models \citep{1993MNRAS.265..250D, 1994AJ....107..634T}, and the Einasto models \citep{1965TrAlm...5...87E, 2022A&A...667A..47B}. Dynamical systems, including galaxies and star clusters, possess finite spatial dimensions, prompting exploration into the feasibility of constructing models reflecting this characteristic. A prevalent approach involves implementing a truncation based on binding energy within the distribution function. This method selectively excludes orbits with the lowest binding energies, effectively limiting the inclusion of particles or stars beyond a specified truncation radius.

Nevertheless, this approach presents two primary concerns. Firstly, truncating based on binding energy introduces an artificial constraint on the system, eliminating the population of certain orbits within the model despite their gravitational binding and confinement within the permissible radial range. Notably, nearly circular orbits near the system's truncation radius are excluded when energy truncation is applied \citep{1988ApJ...325..566K}. Secondly, a truncation of a given distribution function in binding energy implies a change in the density and the gravitational potential, with the result that one precludes the upfront specification of the density profile. Consequently, crucial dynamical properties typically require numerical computation. The most famous example is the family of King models: they are constructed by truncating the isothermal sphere in binding energy, but important quantities such as the density, potential, and velocity dispersion profiles do not have closed analytical expressions \citep{1966AJ.....71...64K,  2008gady.book.....B}.

We started an investigation into the question whether it is possible to set up self-consistent dynamical models for spherical systems with a preset, analytical density profile, with a finite extent, and ideally, supported by a various orbital structures. In \citetalias{2022MNRAS.512.2266B} \citep{2022MNRAS.512.2266B}, our investigation commenced by scrutinising the uniform density sphere, the most basic among all models with a finite extent \citep[see also][]{Zeldovich1972, 1979PAZh....5...77O}. We presented a series of self-consistent dynamical models for the uniform density sphere, wherein all feasible orbits are fully populated. In \citetalias{2023MNRAS.519.6065B} \citep{2023MNRAS.519.6065B} we demonstrated that no spherical models with a discontinuous density truncation can be supported by an isotropic orbital structure. On the other hand, we showed that many radially truncated models can be supported by an Osipkov--Merritt orbital structure that becomes completely tangential at the truncation radius. In \citetalias{2023MNRAS.525.1795B} \citep{2023MNRAS.525.1795B} we investigated the dynamical properties of a family of power-law spheres with a general tangential Cuddeford orbital structure. 

These papers have enlarged the available suite of self-consistent dynamical models with a finite extent and with analytical dynamical properties. One limiting characteristic of these models, however, is that they all have a dynamical structure that systematically grows more tangential towards the outskirts, with complete tangential anisotropy at the truncation radius ($\rT$). In the present paper, we go one step further and we aim at constructing dynamical models with a more diverse range of orbital structures, including isotropic models. From the analysis in \citetalias{2023MNRAS.519.6065B} we know that isotropic models with a finite extent are only possible for density profiles without a discontinuity. In other words, we need to concentrate on models in which the density smoothly tends to zero at $r=\rT$. 

A continuity at the truncation radius might not be sufficient, however. Our dynamical analysis of the family of broken power-law model \citep[BPL:][]{2020ApJ...892...62D} suggests that models in which the first derivative of the density is discontinuous also pose problems. Indeed, we found that none of the BPL models, which show a discontinuity in the first derivative of the density, has an isotropic distribution function that is positive over the entire phase space, implying that the BPL models cannot be supported by an isotropic orbital structure  \citep{2021MNRAS.503.2955B}. This strongly suggests that we should focus on models that show a sufficiently smooth transition at the break radius.

In this paper we present a family of models that satisfy these requirements. The models are based on Wendland functions, the most popular suite of radial basis functions with a finite support. In Section~{\ref{Construction.sec}} we present the Wendland functions and the construction of the dynamical models based on them, which we call Wendland models. In Section~{\ref{Basic.sec}} we present the basic properties of these models, calculated using the {\tt{SpheCow}} code. In Section~{\ref{Isotropic.sec}} we demonstrate that all the Wendland models can be supported by an isotropic distribution function and we discuss the most important dynamical properties. In Section~{\ref{OM.sec}} we discuss Wendland models with an general Osipkov--Merritt orbital structure. We show that the Wendland models can be supported by a range of Osipkov--Merritt distribution functions, both radially and tangentially anisotropic. In Section~{\ref{LP.sec}} we show that we can extend our suite of self-consistent models by simple linear superposition, which allows to generate models with an even wider range of orbital properties. We discuss and summarise our results in Section~{\ref{Conclusion.sec}}.

\section{Construction of the Wendland models}
\label{Construction.sec}

\subsection{Compactly supported radial basis functions}

Inspiration to build up dynamical models with a finite extent was found in the field of smoothed particle hydrodynamics (SPH) \citep{1977AJ.....82.1013L, 1977MNRAS.181..375G, 2010ARA&A..48..391S}. The essence of SPH is that all physical quantities are smoothed in space through a convolution with a smoothing kernel. The obvious choice for a smoothing kernel is a Gaussian function, but it was quickly realised that smoothing kernels with a finite support have clear advantages. The most popular smoothing kernels used in SPH nowadays are spline kernels , such as the $M_3$ spline kernel \citep{1981csup.book.....H} and the $M_4$ spline kernel \citep{1985A&A...149..135M}. 

The kernels used in SPH belong to a broader class of functions called radial basis functions. They are a critical ingredient for the interpolation of scattered data in multiple dimensions, with ample applications ranging from neural networks and machine learning to the solution of differential equations in high dimensions \citep[for a general overview, see][]{Wendland2005, Fasshauer2007}. In order to be useful for modelling scattered data, radial basis functions need to be positive definite. As in the field of SPH, the property of finite support has a number of major advantages for radial basis functions. The search for suitable compactly supported radial basis functions and their efficient calculation is the topic of extensive research \citep[e.g.,][]{Wu1995, Wu2005, Buhmann1998, Buhmann2000, Gneiting1999, Li2015, Argaez2017, Menandro2019, Bjornsson2019, Bjornsson2021, Hubbert2023}. 

The most commonly used family of compactly supported radial basis functions is the one presented by \citet{Wendland1995}, and which has become known as the family of Wendland functions.\footnote{They are sometimes denoted as the {\em{original}} Wendland functions. This family has been extended by \citet{Schaback2011} with a set of functions, known as the missing Wendland functions.} For a given number of spatial dimensions $d$, the Wendland functions are denoted as $\phi_{\ell,k}(u)$, with the integer number $k$ known as the smoothness parameter and the integer number $\ell$ given by
\begin{equation}
\ell = \left\lfloor \frac{d}{2}+k\right\rfloor + 1.
\end{equation}
All Wendland functions are positive definite and compactly supported on the interval $[0,1]$. The interesting property of the Wendland functions is that they are simple polynomials in $u$ of order $2k+\ell$. More specifically, the $k$'th Wendland function is the unique compactly supported polynomial function with smoothness $k$ that has the lowest possible polynomial degree \citep{Wendland1995}. In the 3D case, with $\ell=k+2$, one has
\begin{gather}
\phi_{2,0}(u) = (1-u)_+^2,
\label{phi20}
\\
\phi_{3,1}(u) = \frac{1}{20}\,(1-u)_+^4\,(1+4u)
\\[0.3em]
\phi_{4,2}(u) = \frac{1}{1680}\,(1-u)_+^6\,(3+18u+35u^2)
\\[0.3em]
\phi_{5,3}(u) = \frac{1}{22176}\,(1-u)_+^8\,(1+8u+25u^2+32u^3),
\\[0.3em]
\phi_{6,4}(u) = \frac{1}{5765760}\,(1-u)_+^{10}\,(5+50u+210u^2+450u^3+429u^4).
\end{gather}
In these expressions, $x_+^\alpha$ represents the truncated power function \citep{Massopust2010},
\begin{equation}
x^\alpha_+ = x^\alpha\,\Theta(x) 
=
\begin{cases} \; 0 & \quad{\text{for }}x\leqslant 0, \\
\;x^\alpha & \quad{\text{for }}x> 0.
\end{cases}
\end{equation}
There are various general expressions that can be used to calculate the Wendland functions for arbitrary values of the parameters $k$ and $\ell$ \citep{Hubbert2012, Chernih2014a, Chernih2014b, Argaez2017}. It is also possible to generalise the standard Wendland functions to non-integer values of the parameters \citep{Chernih2014a}. In the frame of the current work, generalised Wendland functions with non-integer parameters have little added value.
 
\subsection{Construction of the Wendland models}

\begin{table*}
\caption{The dimensionless parameter $q_k$, the central potential well $\Psi_0$, the total potential energy $W_{\text{tot}}$, the boundaries of the allowed $\lambda$ values, and the maximum anisotropy as a function of the smoothness parameter $k$ for a number of Wendland functions. All quantities are expressed in dimensionless units with $G=M_{\text{tot}} = r_{\text{h}} = 1$.}
\label{rhoW.tab}
\centering
\begin{tabular}{ccccccc}
\hline\hline\\[-0.5em]
$k$ & $q_k$ & $\Psi_0$ & $W_{\text{tot}}$ & $\lambda_{\text{min}}$ & $\lambda_{\text{max}}$ & $\lambda_{\text{max}}$\\[0.5em]
\hline \\
0 & 2.00000 & $1.25000$ & $-0.44643$ & $-0.25000$ & 0.97427 & 0.79580 \\
1 & 2.43431 & $1.23238$ & $-0.44124$ & $-0.16875$ & 0.87974 & 0.83905 \\
2 & 2.79912 & $1.22807$ & $-0.43914$ & $-0.12763$ & 0.72764 & 0.85077 \\
3 & 3.12075 & $1.22667$ & $-0.43800$ & $-0.10268$ & 0.63505 & 0.86082 \\ 
4 & 3.41194 & $1.22616$ & $-0.43729$ & $-0.08590$ & 0.60830 & 0.87626 \\
5 & 3.68006 & $1.22599$ & $-0.43679$ & $-0.07384$ & 0.60147 & 0.89066 \\
6 & 3.92991 & $1.22595$ & $-0.43643$ & $-0.06475$ & 0.60059 & 0.90268 \\
7 & 4.16479 & $1.22597$ & $-0.43615$ & $-0.05765$ & 0.60178 & 0.91258 \\
8 & 4.38711 & $1.22601$ & $-0.43592$ & $-0.05196$ & 0.60371 & 0.92076 \\
$\infty$ & $\infty$ & $1.22728$ & $-0.43391$ & 0 & 0.64485 & $\infty$ \\[0.5em]
\hline\hline
\end{tabular}
\end{table*}

\begin{figure}
\includegraphics[width=0.95\columnwidth]{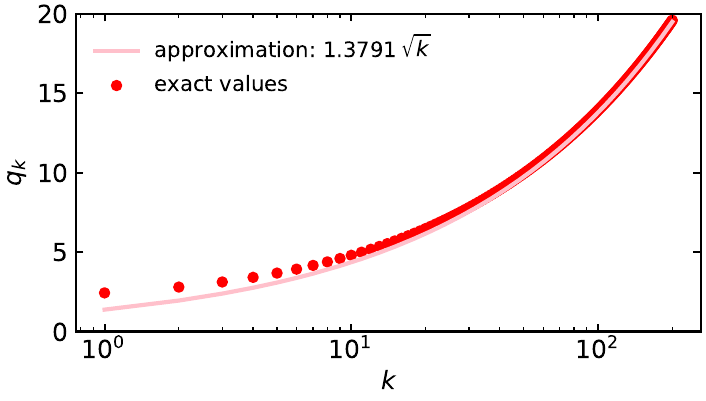}\hspace*{2em}%
\caption{The dimensionless parameter $q_k$ as a function of the smoothness parameter $k$. The red dots show the exact value, the pink line corresponds to the first-order asymptotic approximation $q_k \approx 1.3791\sqrt{k}$.}
\label{Wendland-q.fig}
\end{figure}

The Wendland functions are an interesting starting point to generate a family of self-consistent spherical dynamical models with a finite support. For every integer value of the smoothness parameter $k$, we can define the density profile
\begin{equation}
\rho(r) 
= 
\frac{(3k+5)!}{8\pi\,2^k\,(k+1)!\,(k+2)!}\,
\frac{M_{\text{tot}}}{q_k^3r_{\text{h}}^3}\,\phi_{k+2,k}\left(\frac{r}{q_k r_{\text{h}}}\right).
\label{rhoW}
\end{equation}
We call this family of models the Wendland models. Apart from the smoothness parameter $k$, the expression (\ref{rhoW}) contains two free parameters: the total mass $M_{\text{tot}}$ and the half-mass radius $r_{\text{h}}$. The first factor in the expression above is introduced to ensure a proper normalisation \citep{Chernih2014b}. The parameter $q_k$ in the expression above is a dimensionless parameter introduced to ensure that $r_{\text{h}}$ is the actual half-mass radius. It can be found as the solution of the equation 
\begin{equation}
4\pi \int_0^{r_{\text{h}}} \rho(r)\,r^2\,\txd r = \frac{M_{\text{tot}}}{2},
\end{equation}
or equivalently,
\begin{equation}
\int_0^{1/q_k} \phi_{k+2,k}(u)\,u^2\,\txd u = \frac{2^k\,(k+1)!\,(k+2)!}{(3k+5)!}.
\label{findq}
\end{equation}
For each value of $k$ this is a polynomial equation in $1/q_k$. Combining expressions~(\ref{phi20}) and (\ref{findq}) it is easy to recover the value $q_0=2$ for $k=0$. For other values of $k$, the number $q_k$ must be determined numerically. A number of values are listed in the second column of Table~{\ref{rhoW.tab}}, where it is seen that the value of $q_k$ increases with increasing $k$. Using an asymptotic analysis (see later), one can find that $q_k \propto \sqrt{k}$ for $k\gg1$. This asymptotic behaviour is illustrated in Fig.~{\ref{Wendland-q.fig}}. The value of $q_k$ is an important characteristic of the models: since the Wendland functions are identically zero for arguments larger than one, the models defined by the density profile (\ref{rhoW}) have a truncation radius
\begin{equation}
\rT = q_k r_{\text{h}}.
\end{equation} 

\section{Basic properties}
\label{Basic.sec}

\begin{figure*}
\includegraphics[width=\textwidth]{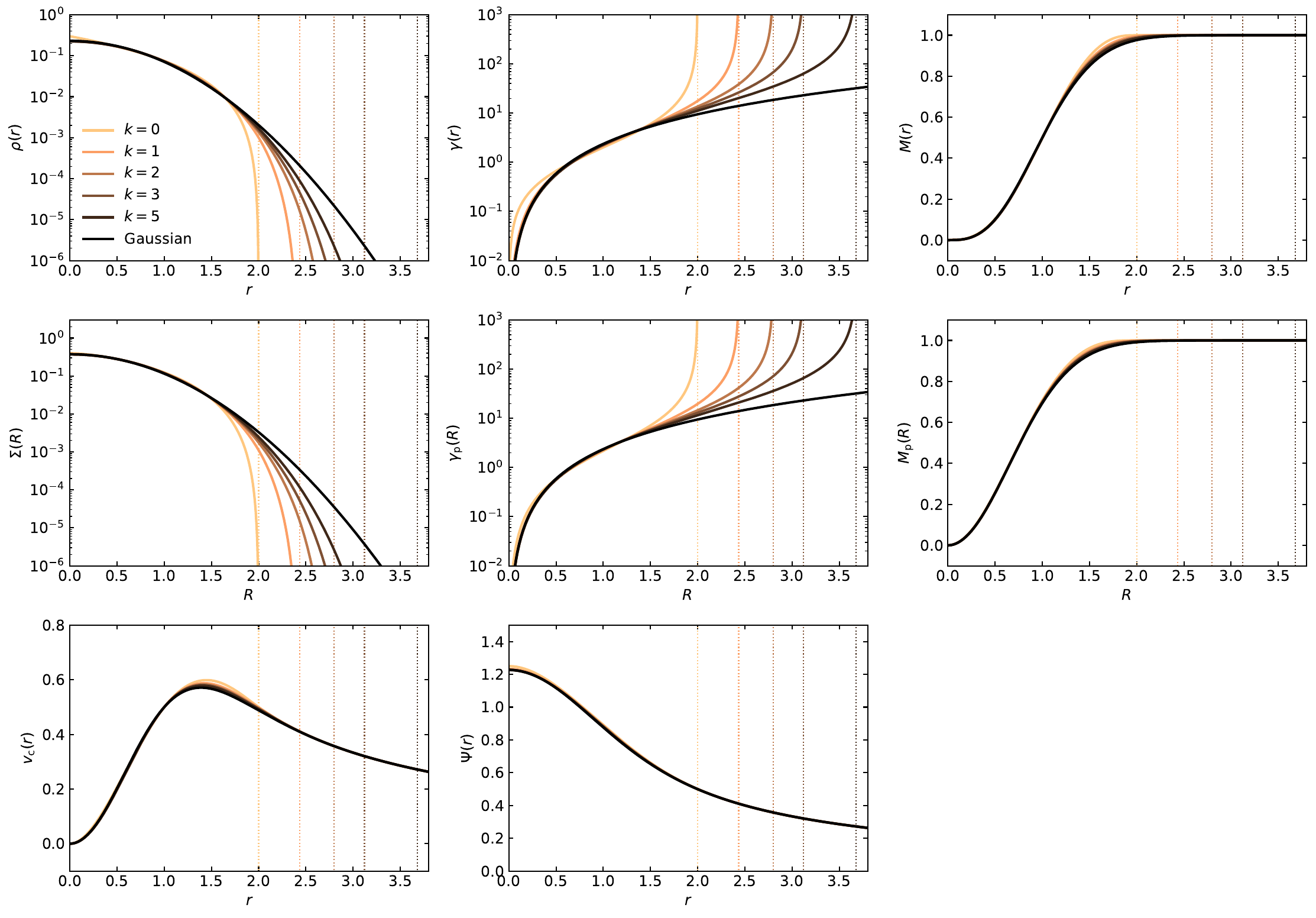}%
\caption{Basic properties of the family of Wendland models defined by the density profile~(\ref{rhoW}). The different colours correspond to different values of the Wendland smoothness parameter $k$ as indicated in the top left panel. The dotted vertical lines in the different panels indicate the truncation radius $\rT = q_k r_{\text{h}}$. We have used dimensionless units with $G=M_{\text{tot}}=r_{\text{h}}=1$.}
\label{Wendland-basic.fig}
\end{figure*}

We now consider some basic properties of the family of Wendland models. Since the density profile is a polynomial function of radius for all models, most basic properties can be analytically. In particular, for each individual Wendland model, the cumulative mass and the gravitational potential are polynomial functions of radius as well, and the surface density and cumulative surface mass profile can be written in terms of elementary functions. However, these analytical expressions quickly become very long as $k$ increases. 

We have therefore implemented the Wendland models in the {\tt{SpheCow}} code \citep{2021A&A...652A..36B}. This code is specifically designed to investigate the structural properties and dynamics of any spherical model defined by an analytical density profile or surface density profile. The only requirement is an implementation of the density and its first and second order derivatives; the code calculates all important structural and dynamical properties automatically, based on high-order Gauss-Legendre integration schemes. Previous work using {\tt{SpheCow}} includes detailed studies of the families of broken power-law, S\'ersic, Nuker and Einasto models \citep{2019A&A...626A.110B, 2020A&A...634A.109B, 2022A&A...667A..47B, 2021MNRAS.503.2955B}. The Wendland models, defined by analytical density profiles, are ideally suited to be explored with {\tt{SpheCow}}.

Fig.~{\ref{Wendland-basic.fig}} shows the most important basic properties of a number of Wendland models with different smoothness parameter $k$. The first two panels of the top row show the density profiles and the logarithmic density slope profiles, defined through
\begin{equation}
\gamma(r) = -\frac{\txd\ln\rho}{\txd\ln r}(r).
\end{equation}
These panels show that the Wendland models have a finite extent. The truncation radius, which increases with increasing $k$, is indicated as the dotted vertical line in each panel. In the inner regions ($r\lesssim1.5\,r_{\text{h}}$), the density of all Wendland models is very similar. The only exception is the $k=0$ model that has a slightly different behaviour at the very centre: it has a non-zero derivative of the density at the centre, whereas all other Wendland models have a zero derivative. 

Also shown on these panels is the Gaussian model with the same mass and half-mass radius,
\begin{equation}
\rho(r) = \frac{1}{\pi^{3/2}}\, \frac{M_{\text{tot}}}{\xi^3 r_{\text{h}}^3} \exp\left[-\left(\frac{r}{\xi\,r_{\text{h}}}\right)^2\right],
\end{equation}
with
\begin{equation}
\xi = 0.919412
\end{equation}
a dimensionless parameter that guarantees that $r_{\text{h}}$ corresponds to the half-mass radius. \citet{Chernih2014b} formally demonstrated that the family of Wendland radial basis functions, when properly renormalised, uniformly converges to a Gaussian function for $k\to\infty$. We see that, indeed, the density of the family of Wendland models converges to the density profile of a Gaussian model. Contrary to the Wendland models, the Gaussian model is not truncated and has an infinite support. This is in agreement with the fact that $q_k$ increases asymptotically as $\sqrt{k}$: in the limit $k\to\infty$ we have $\rT = q_k r_{\text{h}}\to\infty$. 

The first two panels of the second row show the surface density profile and its logarithmic slope,
\begin{equation}
\gamma_{\text{p}}(R) = -\frac{\txd\ln\Sigma}{\txd\ln R}(R),
\end{equation}
for same set of Wendland models. The surface brightness profiles are very similar to the density profiles, with the same systematic behaviour as a function of $k$. The surface density profile is almost independent of $k$ in the inner regions, and we only see differentiation between the different models in the outer regions. Even for the $k=0$ model, the inner part of surface density is almost indistinguishable from the models with larger values of $k$, and even from the Gaussian model.

The remaining panels show a number of other basic properties: the cumulative mass profile, the cumulative projected mass, the circular velocity curve, and the gravitational potential. All of these profiles have a very weak dependence on the smoothness parameter. Given this similarity, it is interesting that there is still so much diversity in the density and surface density profiles. 

All Wendland models have a finite potential well, for which we find the closed expression
\begin{equation}
\Psi_0 = \frac{(3k+5)\,(2k+2)!}{2^{2k+2}\,[(k+1)!]^2}\,\frac{1}{q_k}\,\frac{GM_{\text{tot}}}{r_{\text{h}}}.
\label{Psi0}
\end{equation}
This expression depends very weakly on $k$, as shown in the third column of Table~{\ref{rhoW.tab}}. It can be used to derive an asymptotic expression for the parameter $q_k$ in the limit $k\gg1$. Indeed, in this limit, expression (\ref{Psi0}) has the asymptotic expansion
\begin{equation}
\Psi_0 \approx 
\frac{1}{\sqrt\pi}\left(3\sqrt{k} + \frac{25}{8}\,\frac{1}{\sqrt{k}} + \cdots\right)
\frac{1}{q_k}\,\frac{GM_{\text{tot}}}{r_{\text{h}}}. 
\end{equation}
On the other hand, the Gaussian model has as central potential 
\begin{equation}
\Psi_0 = \frac{2}{\sqrt\pi}\,\frac{1}{\xi}\,\frac{GM_{\text{tot}}}{r_{\text{h}}}. 
\end{equation}
Combining these two expressions, we have 
\begin{equation}
q_k \approx \xi \left(\frac32\sqrt{k} + \frac{25}{16}\,\frac{1}{\sqrt{k}} + \ldots\right).
\end{equation}
The solid pink line in Fig.~{\ref{Wendland-q.fig}} shows the first term in this approximation.

\section{Isotropic dynamical models}
\label{Isotropic.sec}

\begin{figure*}
\centering
\includegraphics[width=0.66\textwidth]{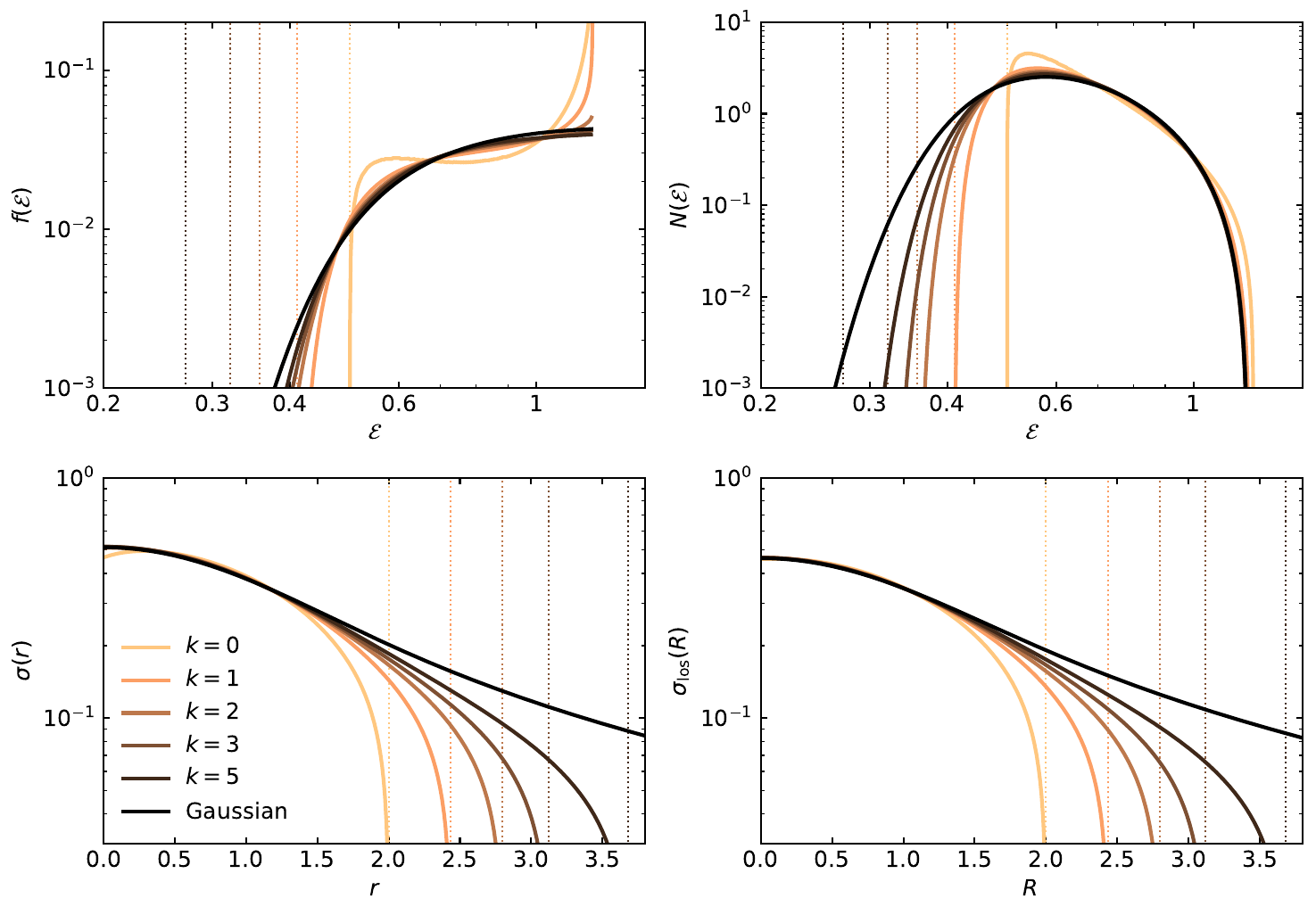}%
\caption{Dynamical properties of the family of Wendland models under the assumption of an isotropic dynamical structure. The dotted vertical lines in the panels on the top row indicate the truncation binding energy $\calET = \Psi(\rT)$; in the bottom row panels the dotted lines correspond to the truncation radius $\rT$. The color scheme is identical to Fig.~{\ref{Wendland-basic.fig}}. We have used dimensionless units with $G=M_{\text{tot}}=r_{\text{h}}=1$.}
\label{Wendland-iso.fig}
\end{figure*}

One of the main goals of this paper is investigating whether it is possible to construct consistent dynamical models with a given density profile with a finite extent. An isotropic dynamical structure is the most obvious first option to explore. Fig.~{\ref{Wendland-iso.fig}} show the most important dynamical properties of the family of Wendland models under the assumption of an isotropic orbital structure. 

\subsection{Distribution function}

For any spherical density profile and associated potential, we can calculate the unique ergodic or isotropic distribution function by simply apply applying Eddington's formula \citep{2008gady.book.....B, 2021isd..book.....C}. The resulting distribution function $f(\calE)$ is a function of binding energy alone. Whether or not this corresponds to a viable dynamical model depends on whether $f(\calE)$ is nonnegative for all binding energies. 

The top left panel of Fig.~{\ref{Wendland-iso.fig}} shows the distribution function for the same suite of Wendland models as considered in Fig.~{\ref{Wendland-basic.fig}}. Since each model is characterised by a finite extent, the distribution function is zero for all $\calE<\calET$, with 
\begin{equation}
\calET = \Psi(\rT) = \frac{GM_{\text{tot}}}{\rT} = \frac{GM_{\text{tot}}}{q_k r_{\text{h}}}. 
\end{equation}
As all Wendland models also have a finite potential well, the range of binding energies that can be populated is limited to $\calET \leqslant \calE \leqslant \Psi_0$. We see that the distribution function for all Wendland models is positive within this range of binding energies, implying that all Wendland models can be supported by an isotropic orbital structure.

Interestingly, the distribution function of the different models seems to show a different behaviour. For the $k=0$ model it increases very steeply for $\calE \gtrsim \calET$, subsequently it decreases as a function of $\calE$, and finally it diverges when $\calE$ approaches $\Psi_0$. As the distribution function is not monotonically increasing over the entire range of binding energies, it is not guaranteed that this model is stable against radial and non-radial perturbations. The distribution function of the $k=1$ model increases more gently for $\calE \gtrsim \calET$, and is a monotonically increasing function of binding energy over the entire range. This distribution function also diverges when the binding energy approaches the maximum value $\Psi_0$. Finally, the models with $k\geqslant2$ all have a qualitatively similar distribution function that increases smoothly and monotonically from zero at $\calE = \calET$ to a finite value at $\calE = \Psi_0$. In the limit $k\to\infty$, the Gaussian model, we find the same behaviour, but now with $\calET \to 0$. This finding is in agreement with our previous studies of the families of S\'ersic and Einasto models: the Gaussian model belongs to both of these families and represents the most concentrated model in both families that still allows a positive ergodic distribution function \citep{2019A&A...626A.110B, 2022A&A...667A..47B}. The significant differences between the shape of the distribution functions of the Wendland models with $k=0$, $k=1$ and $k\geqslant2$ are remarkable, given the very similar shape of the density profile, and particularly of the mass profile and gravitational potential. 

\subsection{Differential energy distribution}

The shape of the function $f(\calE)$ needs to be interpreted with the necessary caution: a diverging distribution function does not imply a diverging number of stars or particles with large binding energies. In this respect, the differential energy distribution, shown in the top right panel of Fig.~{\ref{Wendland-iso.fig}}, gives more physical insight. We checked the accuracy of our {\tt{SpheCow}} calculations  by integrating the differential energy distribution over the entire energy range and recovered the total mass, as required. 

All Wendland models have, qualitatively, a similar differential energy distribution that converges to zero both in the low ($\calE\to\calET$) and the high binding energy ($\calE\to\Psi_0$) limits. For all models, the peak in the differential energy distribution lies between 0.5 and $0.6\times GM_{\text{tot}}/r_{\text{h}}$. The mean value of the binding energy, or equivalently, the total integrated binding energy $B_{\text{tot}} = M_{\text{tot}} \langle\calE\rangle$, satisfies the fundamental energy relation,
\begin{equation}
B_{\text{tot}} = 3\,T_{\text{tot}} = -\frac32\,W_{\text{tot}},
\end{equation}
where $T_{\text{tot}}$ represents the total kinetic energy and $W_{\text{tot}}$ the total potential energy \citep{2021A&A...653A.140B}. We find that these energy budgets depend only weakly on the smoothness parameter $k$, as shown in the fourth column of Table~{\ref{rhoW.tab}}.

\subsection{Velocity dispersion profiles}

Finally, the two bottom row panels of Fig.~{\ref{Wendland-iso.fig}} show the intrinsic and the projected velocity dispersion profiles. In all cases, the intrinsic velocity dispersion profile is characterised by a finite central value and a smooth and monotonic decline towards zero at the truncation radius. The only exception is the model with $k=0$ which has a minor depression at the central regions: the velocity dispersion increases in the innermost region, reaches a maximum value at $r \approx 0.288\,r_{\text{h}}$, and subsequently declines smoothly to zero at the truncation radius. 

A very similar behaviour is found for the projected velocity dispersion profile, but with a much smaller difference between the different models in the inner regions.

\section{Models with an Osipkov--Merritt orbital structure}
\label{OM.sec}

\subsection{Type~I and type~II Osipkov--Merritt models}

If an isotropic orbital structure is the number one option for spherical dynamical models, both from a physical and a mathematical point of view, the anisotropic Osipkov--Merritt orbital structure is an interesting number two. The Osipkov--Merritt orbital structure is named after the independent studies by \citet{1979PAZh....5...77O} and \citet{1985AJ.....90.1027M}, but it was originally presented more than 40 years earlier by \citet{1936MNRAS..96..749S}. The defining characteristic of dynamical models with an Osipkov--Merritt orbital structure is that the distribution function is a spheroidal probability distribution in velocity space. \citet{1985AJ.....90.1027M} considered two families of models; radially anisotropic type~I models (Section~{\ref{ROM.sec}}) and tangentially anisotropic type~II models (Section~{\ref{TOM.sec}}).

\subsubsection{Radially anisotropic Osipkov--Merritt models}
\label{ROM.sec}

Type~I models have a distribution function that only depends on $\calE$ and $L$ through the combination 
\begin{equation}
Q = \calE - \frac{L^2}{2r_\txa^2}
=
\Psi(r) - \frac12\left[v_r^2 + \left(1 + \frac{r^2}{r_\txa^2}\,\right)v_{\text{t}}^2\right]
\end{equation}
with $r_\txa$ an arbitrary constant called the anisotropy radius. The anisotropy of the type~I Osipkov--Merritt models is
\begin{equation}
\beta(r) \equiv 1 - \frac{\sigma_{\text{t}}^2(r)}{2\sigma_r^2(r)} = \frac{r^2}{r^2 + r_\txa^2},
\end{equation}
which implies isotropy in the centre and completely radial anisotropy at large radii. A major attractiveness of this class of models is that the distribution function $f=f(Q)$ can be inverted from the density using a simple extension of the Eddington equation that applies to ergodic models. Whether or not the distribution function is positive over the entire phase space, and the model thus consistent, generally depends on the density profile and on the value of $r_\txa$. We encounter three different cases.

On one side of the spectrum we have models that cannot be supported by an isotropic distribution function. Such models cannot be supported by a type~I Osipkov--Merritt orbital structure either, since radially orbital configurations are more demanding than isotropic orbital structures. Examples of such models are the uniform density sphere (\citealt{1979PAZh....5...77O}; \citetalias{2022MNRAS.512.2266B}), S\'ersic models with S\'ersic index $m<\nicefrac12$ \citep{2019A&A...626A.110B}, or the radially truncated Plummer model considered in \citetalias{2023MNRAS.519.6065B}.

On the other side of the spectrum we have models that can be supported by a completely radial orbital structure. These models can also be supported by radial Osipkov--Merritt orbital configurations for any value of the anisotropy radius $r_\txa$. They must have a central density cusp at least as steep at $r^{-2}$ \citep{1984ApJ...286...27R, 2006ApJ...642..752A}. Examples include the $\gamma$--models with $\gamma\geqslant2$ \citep{1995MNRAS.276.1131C}, with the Jaffe model \citep{1985MNRAS.214P..25M} and the $\gamma=\nicefrac52$ model \citep{2004MNRAS.351...18B} as interesting special cases for which the type~I Osipkov--Merritt distribution function can be calculated analytically.

In between these extreme cases are the models that can be supported by an isotropic distribution function but not by a completely radial orbital structure. These models are characterised by a critical value $(r_\txa)_{\text{c}}$ that forms the boundary between consistent and inconsistent radial Osipkov--Merritt distribution functions. 
Examples include the Plummer model \citep{1979PAZh....5...77O, 1985AJ.....90.1027M}, the Hernquist model \citep{2002A&A...393..485B}, $\gamma$--models with $\gamma<2$ \citep{1995MNRAS.276.1131C}, S\'ersic models with $m>\nicefrac12$ \citep{1997A&A...321..724C}, and Einasto models with $n>\nicefrac12$ \citep{2022A&A...667A..47B}. The Wendland models considered in this paper belong to this third category, so we expect to find, for each value of $k$, a critical anisotropy radius $(r_\txa)_{\text{c}}$ with consistent type~I Osipkov--Merritt models for all $r_\txa$ larger than this value.

\subsubsection{Tangentially anisotropic Osipkov--Merritt models}
\label{TOM.sec}

Type~II Osipkov--Merritt models are characterised by a distribution function that depends on $\calE$ and $L$ through the combination 
\begin{equation}
Q = \calE + \frac{L^2}{2r_\txa^2}
=
\Psi(r) - \frac12\left[v_r^2 + \left(1 - \frac{r^2}{r_\txa^2}\,\right)v_{\text{t}}^2\right]
\end{equation}
The distribution function can be inverted from the density using a very similar inversion formula as in the case of type~I models. For type~II models, the anisotropy profile is
\begin{equation}
\beta(r) = \frac{r^2}{r^2 - r_\txa^2}.
\end{equation}
One major limitation is that the type~II Osipkov--Merritt models is that this approach is only valid for the region $r\leq r_\txa$. An extension applicable for the outer region, $r>r_\txa$ needs to be engineered.  \citet{1985AJ.....90.1027M} provide two possible solutions: either purely circular orbits, or a combination of orbits with varying eccentricities that remain confined to the outer region. In both cases, however, the tangential velocity dispersion is characterised by a discontinuity at $r=r_\txa$, which makes the type~II Osipkov--Merritt orbital structure slightly unphysical. This immediately explains why models with this orbital structure have hardly been explored in the literature.

This problem does not apply to models with a finite extent, however. For models with a truncation radius $\rT$, we can meaningfully define type~II Osipkov--Merritt models with as long as $r_\txa\geq\rT$. In \citetalias{2023MNRAS.519.6065B} we argued that many radially truncated models can be supported by a tangential Osipkov–Merritt orbital structure if $r_\txa=\rT$. A simple example is the uniform density sphere: this model cannot be supported by an isotropic velocity distribution, but the type~II Osipkov--Merritt model with $r_{\text{a}}=\rT$ is consistent and the distribution function has a very simple form (\citealt{1974SvA....17..460P, 1979PAZh....5...77O}; \citetalias{2022MNRAS.512.2266B}). The same applies more widely to models with a truncated power-law density distribution, including the truncated singular isothermal sphere \citepalias{2023MNRAS.525.1795B}.

For models with a finite extent that can be supported by an isotropic distribution function, we expect that all type~II Osipkov--Merritt models with $r_\txa \geqslant \rT$ are consistent, since populating a model with tangential orbits is less demanding than populating it with a mixture of orbits that combines to orbital isotropy. This particularly applies to the Wendland models discussed in this paper.

\subsubsection{A unified family}

Summarising the analysis from the two previous subsections, it can be expected that, for models with a finite extent that can be supported by an isotropic orbital structure, we can find a range of both type~I and type~II Osipkov--Merritt distribution functions that are positive over the entire phase space. An elegant way express this is to use the notation adopted by \citet{1979PAZh....5...77O}. Instead of considering two families, each with an anisotropy radius $r_\txa$, he unified the type~I and type~II models as a single continuous sequence with $\lambda$ as a free parameter, where 
\begin{gather}
Q = \calE - \tfrac12\,\lambda\,L^2,
\\
\beta(r) = \frac{\lambda\,r^2}{1+\lambda\,r^2},
\label{anisotropylambda}
\\
r_\txa = |\lambda|^{-1/2}.
\end{gather}
Isotropic models are characterised by $\lambda=0$, radially anisotropic type~I models have $\lambda>0$, and tangentially anisotropic models have $\lambda<0$. 

\subsection{Wendland Osipkov--Merritt models}

\begin{figure*}
\newlength{\imageheight}
\settoheight{\imageheight}{\includegraphics{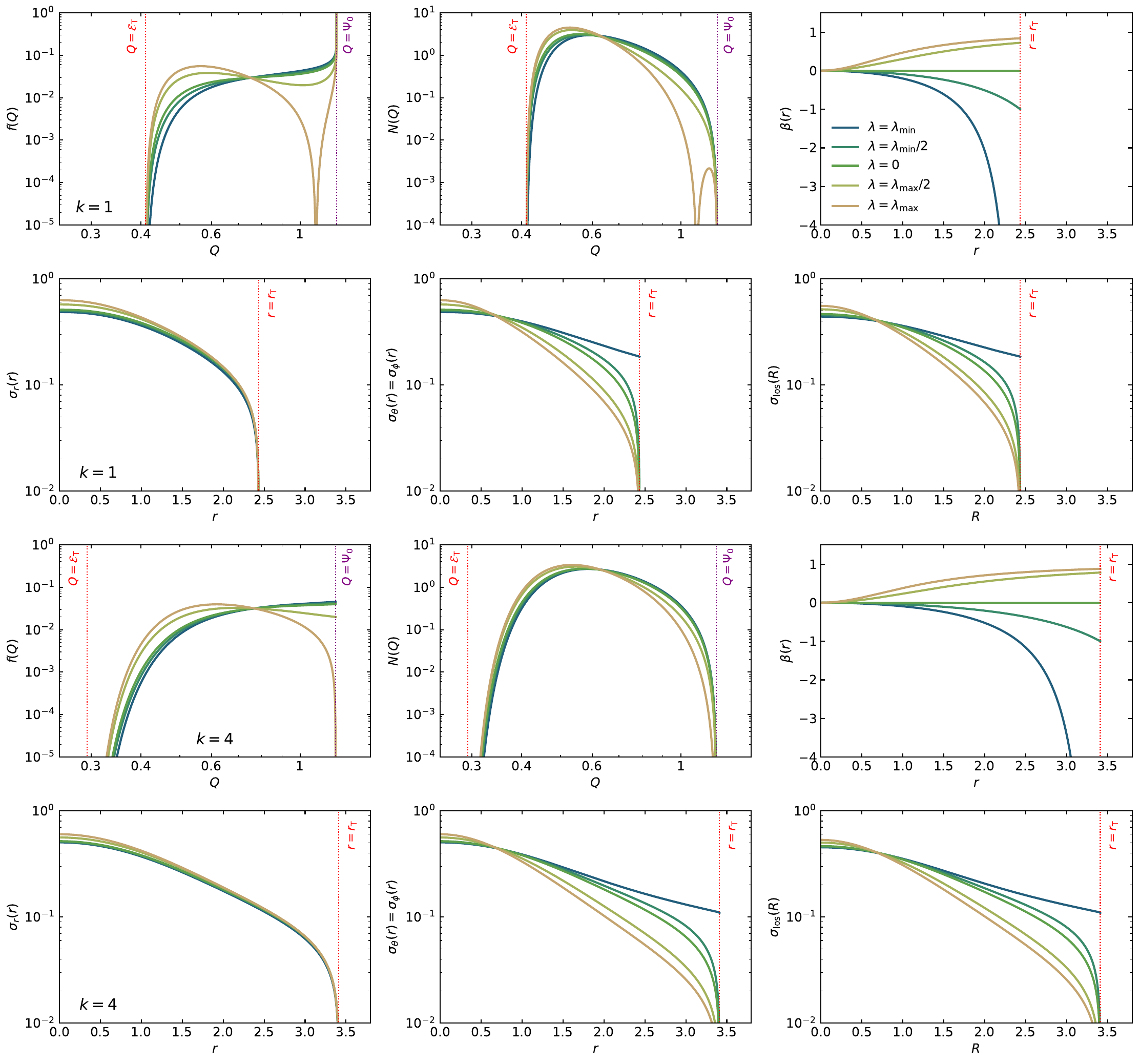}} 
\includegraphics[trim=0 0.5\imageheight{} 0 0, clip, width=\textwidth]{Wendland-OM.pdf}\\[1.5em]
\includegraphics[trim=0 0 0 0.5\imageheight{}, clip, width=\textwidth]{Wendland-OM.pdf}%
\caption{Dynamical properties of two selected members of the family of Wendland models ($k=1$ on the first two rows, and $k=4$ on the last two rows) under the assumption of an Osipkov--Merritt dynamical structure. The different lines correspond to models ranging from an extremely tangential to an extremely radial orbital structure. We have used dimensionless units with $G=M_{\text{tot}}=r_{\text{h}}=1$.}
\label{Wendland-OM.fig}
\end{figure*}

\subsubsection{Distribution function $f(Q)$}

Based on the discussion above we expect, for every member of the family of Wendland models, a range of consistent models for $\lambda$ in a compact interval $[\lambda_{\text{min}},\lambda_{\text{max}}]$, with
\begin{gather}
\lambda_{\text{min}} = -\frac{1}{\rT^2} = -\frac{1}{q_k^2 r_{\text{h}}^2} < 0,
\\
\lambda_{\text{max}} = \lambda_{\text{max}}(k) > 0.
\end{gather}
We used the {\tt{SpheCow}} code to check this assumption and to numerically determine the boundary $\lambda_{\text{max}}(k)$  for the lowest-order Wendland models. The results are presented in the fifth and sixth columns of Table~{\ref{rhoW.tab}}. The value of $\lambda_{\text{max}}$ is largest for $k=0$, and gradually decreases for increasing $k$, until it reaches a minimum values for $k=6$. It subsequently increases very weakly and reaches the value $\lambda_{\text{max}} = 0.64485\,r_{\text{h}}^{-2}$ for the Gaussian model. This consistency limit is equivalent to the critical anisotropy radius $(r_\txa)_{\text{c}} = 1.24529\,r_{\text{h}}$ obtained by \citet{2022A&A...667A..47B}.

The left panels on the first and third row of Fig.~{\ref{Wendland-OM.fig}} show the distribution function for a set of Wendland models for $k=1$ and $k=4$. For each of the two models, we show $f(Q)$ for five different orbital structures: the most extreme tangential model ($\lambda=\lambda_{\text{min}}$), a moderately tangential model ($\lambda=\tfrac12\,\lambda_{\text{min}}$), the isotropic model ($\lambda=0$), a moderately radial models ($\lambda=\tfrac12\,\lambda_{\text{max}}$) and the limiting radial model ($\lambda=\lambda_{\text{max}}$).

All distribution functions for the $k=1$ model (top left panel)  diverge for $Q\to\Psi_0$. This is to be expected, as Osipkov--Merritt models are isotropic in their central regions and the isotropic $k=1$ model also diverges for $\calE\to\Psi_0$ (see leftmost panel of Fig.~{\ref{Wendland-iso.fig}}). The shape of the $f(Q)$ curve gradually changes as $\lambda$ varies from negative to positive values. The tangential models have a distribution function that is monotonically increasing as a function of $Q$. As the tangential anisotropy decreases, the distribution shifts towards smaller $Q$. This is the result of the gradual replacement of circular-like orbits in the outer regions by more eccentric orbits. This trend continues when $\lambda$ becomes positive. For moderately radially anisotropic models, such as the one with $\lambda = \tfrac12\,\lambda_{\text{max}}$, the distribution function is no longer monotonically increasing, but shows a local minimum at intermediate $Q$ values. For the most radially anisotropic Osipkov--Merritt model that is still consistent, the distribution function shows a strong depression and reaches zero at $Q\approx1.1$. Increasing $\lambda$ even further, that is decreasing the anisotropy radius $r_\txa$, would make the distribution function negative. A similar qualitative behaviour for decreasing anisotropy radius is seen for other radially anisotropic Osipkov--Merritt models \citep[e.g.,][]{1985AJ.....90.1027M, 1996ApJ...471...68C, 1997A&A...321..724C, 2022A&A...667A..47B}.

For the $k=4$ model, representative for Wendland models with $k>2$, the picture is slightly different. The distribution functions $f(Q)$ for all consistent type~I and type~II Osipkov--Merritt models converge to a finite value at $Q=\Psi_0$, again in line with the findings for the isotropic models (Fig.~{\ref{Wendland-iso.fig}}). For all tangentially anisotropic models, $f(Q)$ monotonically increases as a function of $Q$. As $\lambda$ increases, we see again a gradual shift from high $Q$ values to lower $Q$ values. In particular, the finite value reached at $Q=\Psi_0$ gradually decreases as $\lambda$ increases. At some point the monotonicity is broken, as illustrated by the $\lambda=\tfrac12\,\lambda_{\text{max}}$ model. The distribution function for the $\lambda=\lambda_{\text{max}}$ model converges to 0 at $Q=\Psi_0$. Increasing $\lambda$ even further would lead to negative values at the highest $Q$ values.

\begin{figure*}
\includegraphics[width=\textwidth]{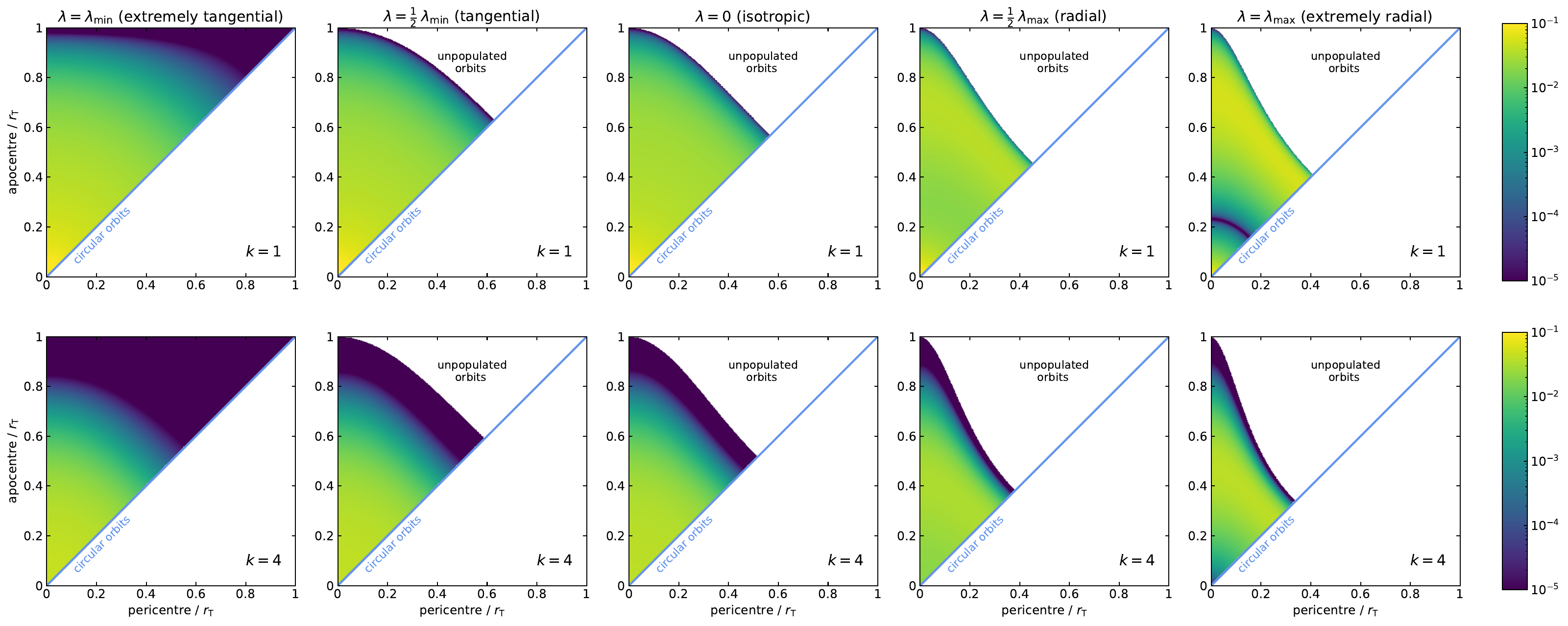}%
\caption{Distribution functions in turning point space for two selected members of the family of Wendland models ($k=1$ on the top row, and $k=4$ on the bottom row) under the assumption of an Osipkov--Merritt dynamical structure. The panels correspond to models ranging from an extremely tangential (left) to an extremely radial orbital structure (right). We have used dimensionless units with $G=M_{\text{tot}}=r_{\text{h}}=1$.}
\label{Wendland-DFTP.fig}
\end{figure*}

\subsubsection{Pseudo-differential energy distribution}

The second panels on the first and third row of Fig.~{\ref{Wendland-OM.fig}} show the the pseudo-differential energy distributions $N(Q)$ of the same set of Osipkov--Merritt models. The quantity $N(Q)$ is the extension of the differential energy distribution for isotropic models and represents the distribution of mass as a function $Q$. It easily can be calculated by multiplying the distribution function by a pseudo-density-of-states function  (\citealt{1991MNRAS.253..414C}, \citetalias{2023MNRAS.519.6065B}). In all cases, $N(Q)$ converges to zero for the lowest and highest $Q$ values, that is, for $Q\to\calET$ and $Q\to\Psi_0$. In both panels, the change of $N(Q)$ with increasing $\lambda$ indicates a shift from low to higher values of $Q$, indicating that tightly bound nearly circular orbits are gradually replaced by more loosely bound eccentric orbits. For the $k=1$ model, the strong depression at intermediate $Q$ that characterises the distribution function for the most radially anisotropic models is translated to a similar depression in the pseudo-differential energy distribution. 

\subsubsection{Distribution function in turning point space}

Comparing the distribution function of different Osipkov--Merritt models written as a function of $Q$ is somewhat contrived, as $Q$ explicitly depends on the anisotropy radius. The quantity $Q$ on the $x$-axis in the different panels of Fig.~{\ref{Wendland-OM.fig}} thus does not represent the same physical quantity for the each of the models plotted in that panel. To compare the different models in an apples-to-apples comparison we plot the distribution function in turning point space in Fig.~{\ref{Wendland-DFTP.fig}}. More specifically, for the Wendland models with $k=1$ (top row) and $k=4$ (bottom row), we show the same five models as shown in Fig.~{\ref{Wendland-OM.fig}}. Comparing the different panels one can note some interesting differences. 

A first obvious difference between the different distribution functions is the coverage of the parameter space. Only the extremely tangential models with $\lambda = \lambda_{\text{min}}$ populates all the possible orbits, that is, all orbits with $0\leqslant r_{\text{peri}} \leqslant r_{\text{apo}} \leqslant \rT$. The larger $\lambda$, the larger the area of turning point space that remains unpopulated. The boundary between the populated and unpopulated area is the line corresponding to $Q = \calET$. These unpopulated orbits are relatively weakly bound, tangential orbits with large apocentres. The existence of a region with orbits that are allowed (they are gravitationally bound and do not cross the truncation radius) but not populated is a natural consequence of the finite extent of the Wendland models. In particular, the same characteristic is observed for all spherical models with a finite extent and an isotropic orbital structure \citepalias{2023MNRAS.519.6065B}. This ``artificial and unphysical'' constraint motivated \citet{1988ApJ...325..566K} to argue against the use of isotropic dynamical models constructed using a truncation in binding energy, such as the popular King models \citep{1966AJ.....71...64K, 2008gady.book.....B}.

A second difference between the different distribution functions is the orientation of the isodensity contours within the populated part of turning-point space. For each model, the distribution function is stratified along curves of constant $Q$, which translates to specific $\lambda$-dependent isodensity contours in turning-point space. At the very centre of each model, these isodensity contours have the same shape for all models, since all Osipkov--Merritt models are isotropic at small radii. At larger radii, however, the isodensity curves have different orientations depending on the value of $\lambda$. For the tangential models, they are relatively horizontal, which indicates the preference for tangential orbits. For radial models, on the other hand, the isodensity contours are more skewed, which indicates the strong preference for radial orbits. 

Finally, apart from the different coverage of parameter space and the orientation of the isodensity contours, the different models differ in how the different $Q$-levels are populated. This is exactly the information contained in the different curves shown in the leftmost panels of Fig.~{\ref{Wendland-OM.fig}}. One notes, for example, the divergence of all distribution functions of the $k=1$ Wendland model at small radii, the depression of the extremely radial Osipkov--Merritt $k=1$ model at intermediate radii ($r\approx0.2\,\rT\approx0.5\,r_{\text{h}}$), and the changing behaviour of the distribution functions of the Wendland $k=4$ model at small radii.

\subsubsection{Velocity dispersion profiles}

Finally, we also briefly discuss the velocity dispersion profiles of the Osipkov--Merritt models for the $k=1$ and $k=4$ models. The panels on the second and fourth row of Fig.~{\ref{Wendland-OM.fig}} show the radial, tangential, and projected velocity dispersion profiles, the rightmost panels of the first and third row show the anisotropy profiles. 

In all cases, the radial dispersion has a finite central value and it gradually decreases towards the truncation radius. The radial dispersion depends only weakly on the orbital structure: it is largest for the most radially anisotropic model and the profile slightly decreases when the anisotropy turns more tangential. The differences are very minor, however. The differences in the tangential velocity dispersion profiles are much more prominent. For the most radial models, the tangential dispersion is highest in the centre and it drops fast when we proceed to the truncation radius. This is due to the prevalence of eccentric orbits populating the outer regions. As the anisotropy becomes more tangential, most stars in the outer regions are on circular-like orbits, and this results in a slower decrease of the projected velocity dispersion profile. In the most extreme tangential model in which all stars on circular orbits in the outskirts of the model, the projected velocity dispersion profile decreases to a finite non-zero value at $r=\rT$. Finally, for the projected velocity dispersion profiles, we find a very similar behaviour as for the tangential dispersions. This is due to the fact that the line-of-sight velocity dispersions are a weighted mean of the radial and tangential dispersions along the line of sight.

The anisotropy profiles have the shape dictated by equation~(\ref{anisotropylambda}): all Osipkov--Merritt models are isotropic in the centre and the anistropy parameter monotonically increases or decreases towards the final value at the truncation radius. Interestingly, even the most radial Osipkov--Merritt model does not reach $\beta=1$ at the truncation radius, but rather the value
\begin{equation}
\beta_{\text{max}} = \frac{\lambda_{\text{max}}\,\rT}{1+\lambda_{\text{max}}\,\rT}.
\end{equation}
The last column in Table~{\ref{rhoW.tab}} lists $\beta_{\text{max}}$ for the different Wendland models. So even for the most radial Osipkov--Merritt models, there are still non-radial, even purely circular, orbits reaching the truncation radius. This can also clearly be seen in the left-most panels of Figure~{\ref{Wendland-DFTP.fig}}.

\section{Linear superposition of models}
\label{LP.sec}

\begin{figure*}
\includegraphics[width=\textwidth]{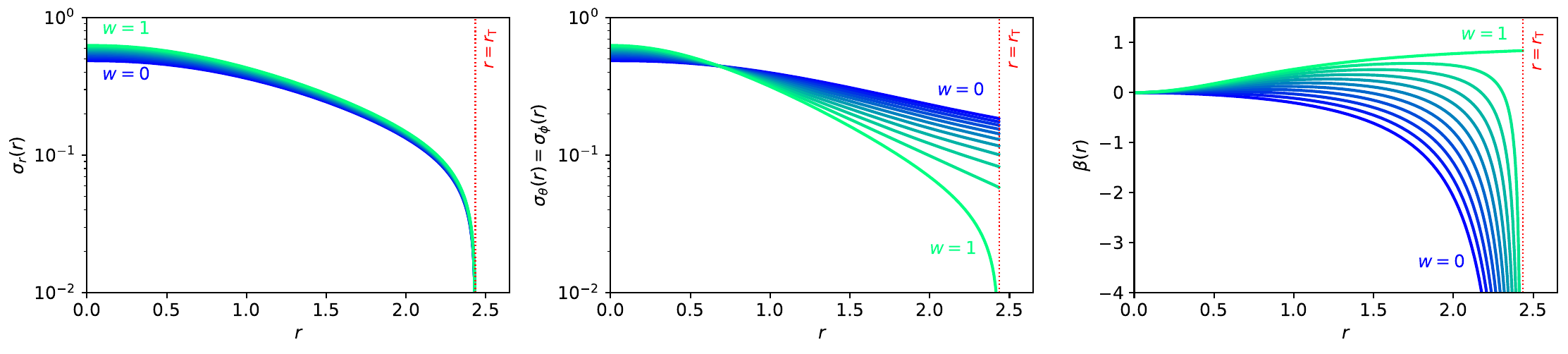}%
\caption{Radial dispersion, tangential dispersion, and anisotropy profiles for the Wendland $k=1$ superposition model described by Equation~(\ref{superp}). The different lines correspond to different values of the weight $w$, varying smoothly between 0 and 1. We have used dimensionless units with $G=M_{\text{tot}}=r_{\text{h}}=1$.}
\label{Wendland-superposition.fig}
\end{figure*}

\begin{figure*}
\newlength{\imageheightb}
\settoheight{\imageheightb}{\includegraphics{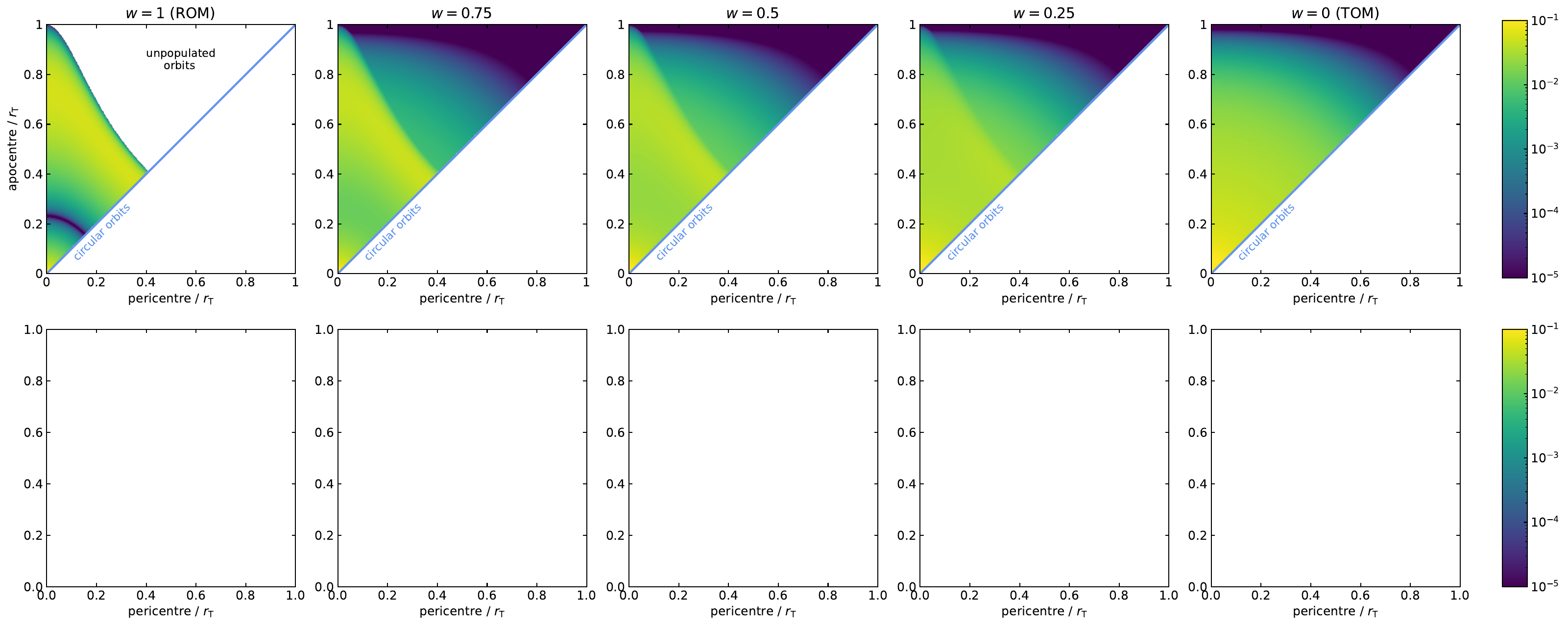}} 
\includegraphics[trim=0 0.5\imageheightb{} 0 0, clip, width=\textwidth]{Wendland-superposition-DFTP.pdf}%
\caption{Distribution functions in turning point space for the Wendland $k=1$ superposition model described by Equation~(\ref{superp}). The different panels correspond to different values of the weight $w$, as indicated on top of each panel. We have used dimensionless units with $G=M_{\text{tot}}=r_{\text{h}}=1$.}
\label{Wendland-superposition-DFTP.fig}
\end{figure*}

The family of Osipkov--Merritt models presented in the previous section allows to generate, for each Wendland density, a suite of models with different anisotropy profiles. We can extend the range of models even further by linear superposition of models \citep[see e.g.,][]{1985AJ.....90.1027M, 1986PhR...133..217D, 1987MNRAS.224...13D}. Indeed, we can, for each value of $k$, consider a linear combination of Osipkov--Merritt models ${\cal{M}}_\lambda$ with different values of $\lambda$ between $\lambda_{\text{min}}(k)$ and $\lambda_{\text{max}}(k)$. A linear superposition can generally be written as
\begin{equation}
{\cal{M}} = \int w(\lambda)\,{\cal{M}}_\lambda\,\txd\lambda 
\quad{\text{with}}\quad 
\int w(\lambda)\,\txd\lambda = 1,
\end{equation}
which reduces to a finite sum in the case of a finite number of models. Since each model ${\cal{M}}_\lambda$ has exactly the same density and the same potential, the new model will obviously also have the same density and potential, and thus every normalised weight function $w(\lambda)$ generates a new model that self-consistently generates the original potential--density pair. In principle, the weight function $w(\lambda)$ can have any functional form, as long as the normalisation condition is fulfilled and the resulting distribution function 
\begin{equation}
F(\calE,L) = \int w(\lambda)\,F_\lambda(\calE,L)\, \txd\lambda
\end{equation}
is nonnegative over the entire phase space. When $w(\lambda) \geq 0$ for all $\lambda$, this condition is trivially satisfied, but with negative weights, this is not necessarily the case.

The anisotropy profile of the new model is easily calculated. Since
\begin{gather}
\rho\sigma_r^2(r) = \int w(\lambda)\,\rho\sigma_{r,\lambda}^2(r)\,\txd\lambda,
\\
\rho\sigma_{\text{t}}^2(r) = \int w(\lambda)\,\rho\sigma_{\text{t},\lambda}^2(r)\,\txd\lambda,
\end{gather}
we obtain
\begin{equation}
\beta(r) 
= 
\int w(\lambda) \left[\frac{\sigma_{r,\lambda}^2(r)}{\sigma_r^2(r)}\right] \beta_\lambda(r)\, \txd\lambda.
\end{equation}
At every radius, the anisotropy of the superposition is thus also a weighted average of the anisotropy of the components, but the weights depend on the radius $r$. 

We illustrate the power of this approach in Figures~{\ref{Wendland-superposition.fig}} and~{\ref{Wendland-superposition-DFTP.fig}}, which show a discrete linear superposition of two $k=1$ Wendland models. We chose the most radial and the most tangential Osipkov--Merritt models, 
\begin{equation}
{\cal{M}} = w\,{\cal{M}}_{\lambda_{\text{max}}} + (1-w)\,{\cal{M}}_{\lambda_{\text{min}}},
\label{superp}
\end{equation}
and consider different values for $w$ in this linear combination. The highest value of $w$ for which the distribution function is everywhere nonnegative is $w=1$, the original most radial Osipkov--Merritt model. This is logical: the radial Osipkov--Merritt model has unpopulated orbits over a large part of turning point space (Figure~{\ref{Wendland-DFTP.fig}}). Any combination with the tangential Osipkov--Merritt model in which the latter has a negative weight will lead to a distribution function that is negative in this part of turning point space. 

The anisotropy for the radial Osipkov--Merritt model increases from 0 in the centre to $\beta_{\text{max}}$ at the truncation radius. When $w$ decreases, this radial model is intermixed with the tangential Osipkov--Merritt model. The result is a model that is isotropic in the centre (all Osipkov--Merritt models are isotropic in the centre, so any linear combination is so too), radially anisotropic over most of the radial range, and tangentially anisotropic in the outskirts. Interestingly, since the most tangential Osipkov--Merritt model fills the entire turning point space with orbits, the same accounts for the linear superposition: all allowed orbits are populated. As $w$ decreases even further, the radial anisotropy become less prominent and the decline towards tangential anisotropy sets in at smaller radii. For $w=0$, we recover the tangential Osipkov--Merritt model for which the anisotropy profile gradually declines to complete tangential anisotropy at the truncation radius. When $w$ decreases even further and becomes negative, the distribution function also becomes negative, so in this case the most radial and tangential Osipkov--Merritt models are the extreme cases of this linear superposition. 

The superposition we have presented here is just a simple illustration. It is clear that, by combining different models, we can cover a larger diversity of orbital structures. 

\section{Discussion and conclusion}
\label{Conclusion.sec}

This work presented in this paper fits in an effort to build self-consistent dynamical models for dynamical systems such as galaxies, clusters, dark matter haloes or globular clusters with a finite radial extent. We specifically aim at models with simple analytical density profiles in which many dynamical properties can be calculated analytically. This turned out to be larger endeavour than initially envisioned. 

In the previous papers in this we have presented different families of models, including the uniform density sphere \citepalias{2022MNRAS.512.2266B}, the truncated Plummer model \citepalias{2023MNRAS.519.6065B}, and the family of truncated power-law spheres \citepalias{2023MNRAS.525.1795B}. In all cases, we have demonstrated that these models cannot be supported by an isotropic orbital structure. On the other hand, we have shown that it is possible to find other orbital structures that self-consistently support these models. More specifically, we considered the tangential Osipkov--Merritt and the tangential Cuddeford orbital structures (the latter being a generalisation of the former). Interestingly, in many of the cases mentioned above, most of the important dynamical properties can be expressed completely analytically, including the distribution function. 

The limitation of these works is that, in order to have dynamical models with a distribution function that is nonnegative over the entire phase space, the orbital structure always had to become gradually more tangential for increasing radius, with pure tangentiality at the truncation radius. This is not really what is observed in real dynamical systems. The most massive elliptical galaxies with cores have an orbital structure that systematically varies from moderately tangential at the centre to moderately radial at large radii \citep{2012ApJ...756..179M, 2014ApJ...782...39T, 2016Natur.532..340T, 2020ApJ...891....4L}. Less massive galaxies display a wide variety of orbital structures \citep{2012MNRAS.423.2177S, 2014MNRAS.439..659N, 2014ApJ...792...59Z, 2018A&A...616A..12G, 2022ApJ...930..153S}. In galaxy clusters and dark matter haloes, the anisotropy is typically found to be isotropic in the centre and mildly radial at larger radii \citep{2001ApJ...563..483T, 2004MNRAS.352..535D, 2011MNRAS.415.3895L, 2012ApJ...752..141L, 2013MNRAS.434.1576W, 2016MNRAS.462..663B, 2021MNRAS.500.3151S}. For globular clusters, orbital structures that becomes very tangential at large radii have been predicted by some simulations \citep{2003MNRAS.340..227B, 2015MNRAS.451.2185S}, whereas detailed modelling of observed globular clusters again yields a more mixed view \citep[e.g.,][]{2006A&A...445..513V, 2021AJ....161...41C}.

In this fourth paper of this series we have taken a step towards models with a larger variety in orbital structures; in particular we aimed at setting dynamical models that are radially truncated, have a simple analytical density profile, and that can be supported by orbital structures that do not necessarily become completely tangential at large radii. In particular, we aimed a setting up models that can be supported by isotropic and even radially anisotropic distribution functions. In \citetalias{2023MNRAS.519.6065B} we already argued that, in order to satisfy this requirement, the density distribution needs to be sufficiently smooth at the truncation radius. A simple continuity instead of a sharp break is not sufficient: in \citet{2021MNRAS.503.2955B} we demonstrated that the family of double power-law models \citep{1996MNRAS.278..488Z} cannot be supported by an isotropic distribution function if the parameter that controls the sharpness of the transition between the inner and outer power-law density profiles is too extreme. 

Armed with these guidelines and inspired by the smoothing functions used in SPH, we set up a family of models based on the family of Wendland models, the most popular class of compactly supported radial basis functions. The models considered, which we call Wendland models, contain, apart from the total mass and the half-mass radius, a single free parameter that controls the smoothness $k$ of the transition at the truncation radius. For each value of $k$, the models have a density profile that is a simple polynomial in $r$. Many other basic dynamical properties, such as the cumulative mass or the gravitational potential, can be expressed as simple polynomials as well. The complexity of these expressions increases for increasing $k$; in the limit the family of Wendland models converges to the Gaussian model, a nice consequence of the family of Wendland models \citep{Chernih2014b}.

Using the {\tt{SpheCow}} tool we investigated different possible orbital structures for the family of Wendland models. First of all, we showed that all Wendland models can be supported by an isotropic distribution function (Section~{\ref{Isotropic.sec}}). While some properties of the family of Wendland models, such as the gravitational potential, depend only very weakly on $k$ (Section~{\ref{Basic.sec}}), we do find a surprising variety in the isotropic distribution function. Indeed, for the $k=0$ and $k=1$ models, the distribution function diverges in the central regions (that is, for $\calE\to\Psi_0$), whereas it converges to a finite value for all other models. The isotropic orbital structure is not the only viable option, however. We showed that, for each value of the smoothing parameter $k$, the Wendland models can be supported by a continuous range of Osipkov--Merritt models, ranging from radially anisotropic to completely tangential anisotropic at the truncation radius (Section~{\ref{OM.sec}}). By means of linear superposition of different Osipkov--Merritt models, it is easy to generate models with an even more diverse orbital structure, including models that are radially anisotropic over most of their radial range, and tangentially anisotropic near the truncation radius (Section~{\ref{LP.sec}}).

The set of models presented in this paper are another step forward towards our goal: self-consistent dynamical models with a simple, pre-set density profile and supported by a range of orbital structures. To the best of our knowledge, the Wendland models presented here are the first family of models for which both radial and tangential Osipkov--Merritt distribution functions have been presented (if we do not count the non-physical type~II Osipkov--Merritt models corresponding to density profiles without radial truncation). Thanks to the combination of analytical simplicity and finite extent, these simple models might be useful for some applications, for example to test numerical orbit integration or dynamical modelling techniques. On the other hand, they do not represent very faithful representations for real dynamical systems, with a density profile that tends towards a Gaussian. However, this study has provided an important insight: models with truncated density profiles can be supported by isotropic and even radially anisotropic orbital structures as long as the truncation is sufficiently smooth. The same principle can be applied to more realistic density profiles and can be the way to construct realistic models with a radial truncation and a suite of compatible orbital structures. Such an investigation is beyond the scope of this paper and might be considered in future work in this series.

\section*{Data availability}

No astronomical data were used in this research. The data generated and the plotting routines will be shared on reasonable request to the corresponding author. The {\tt{SpheCow}} code is publicly available on GitHub.\footnote{\url{https://github.com/mbaes/SpheCow}}

\section*{Acknowledgements}

We thank the referee for useful suggestions that extended the scope of this work. MB acknowledges the financial support from the Flemish Fund for Scientific Research (FWO-Vlaanderen, Senior Research Project G0C4723N) and thanks the members of the UGent SKIRT team for the superb team spirit and the moral support. This work has made use of several Python packages, including {\tt{numpy}} \citep{Harris2020}, {\tt{scipy}} \citep{Virtanan2020}, {\tt{matplotlib}} \citep{Hunter2007}, and {\tt{astropy}} \citep{2013A&A...558A..33A}.

\bibliographystyle{mnras}
\bibliography{mybib}

\bsp	
\label{lastpage}
\end{document}